\documentclass[11pt,twosided]{article}
\usepackage[latin1]{inputenc}
\usepackage{amsmath, graphicx, natbib, layout, verbatim}
\usepackage{setspace}
\usepackage{fancyhdr}
\usepackage{float}
\usepackage{natbib}
\title{Assessment of Aortic Aneurysm Rupture Risk}
\author{
Rafael Izbicki, Ann B. Lee and Ender A. Finol\\\\
\small{Carnegie Mellon University}\\\\
}
\date{May 2011}

\begin{document}
\maketitle 

\abstract{The rupture of an abdominal aortic aneurysm (AAA) is associated with a high mortality.
When an AAA ruptures, 50\% of the patients die before reaching the hospital. Of the patients that are able to reach the
operating room, only 50\% have it successfully repaired \citep{surgery}. Therefore, it is important to find good predictors for immediate risk
of rupture. Clinically, the size of the aneurysm is the variable vascular surgeons usually use
to evaluate this risk. Patients with large aneurysms are often sent to surgery. 
However, many studies have shown that even small aneurysms can rupture and deserve
attention as well \citep{Fillinger}. It is important to find good predictors
of rupture that also avoid unnecessary surgery as all surgeries are associated with possible complications. Here, we use data obtained from 144 computed tomographies of patients from the Western Pennsylvania Allegheny Health System to predict the high risk of rupture
of an aneurysm and also to examine which features are important for this goal.}

\section{Introduction}
Abdominal aortic aneurysm (AAA) is the medical term used when the abdominal aorta becomes extremely large (technically, with a diameter
of more than 3cm). The problem with AAAs is that their rupture is often lethal; many patients
die within minutes. Patients that have AAA can undergo a surgery to have it repaired.
However, this surgery is itself risky, so that only patients that are considered to have a high risk of rupture
are encouraged to undergo it. Most vascular surgeons use the diameter of the aneurysm as the criteria for deciding if
a patient has to undergo surgery. When the diameter is greater than 5.5cm \citep{Shum2010}, the patients
are considered for repair. Patients that already have the aneurysm ruptured or have abdominal pain are also sent to surgery.

The two main goals of this study are:

\begin{enumerate}
\item predict the high risk of rupture of the
abdominal aortic aneurysm
\item understand which covariates are important when predicting the high risk of rupture
\end{enumerate}

In order to do so, we have a data base that consists of patients of two groups: 

\begin{enumerate}
\item"elective repair" (100 patients)
\item "emergent repair" (44 patients)
\end{enumerate}

Elective repair patients are individuals with large aneurysms that were suggested to undergo surgery, even
though they did not display an immediate sign of rupture. On the other hand, patients from the emergent group
were sent to surgery either because they had symptomatic pain or because they already had the aneurysm ruptured.
All patients from this study were treated in Western Pennsylvania Allegheny Health System.

The covariates used in this study were calculated from three-dimensional models that were generated from segmented, contrast-enhanced computed tomography images
\citep{Shum2010}. There are in total 28 covariates,
which include 1-dimensional indices such as diameters and heights, 2-dimensional indices such as measures
of tortuosity and asymmetry, 3-dimensional indices such as the volume and the surface area as well as measurements
of wall thickness at different parts of the aneurysm (a grid was created to do so). The information about thickness was condensed by using
four different statistics: minimum, maximum, average and average thickness at maximum diameter. In the Appendix, one
can find all variables used together with their abbreviations used here. For the formulae and more details of these quantities, see
\citep{Shum2010}. Figure 8 in the Appendix shows some of these features.

The idea of the study is to use this database to evaluate the risk of a patient having his aneurysm ruptured. 
We also want to understand which variables are good predictors of this event. Note that
there is one major assumption here, namely that the variables which are measured do not change when the rupture occurs.
In future work we plan to explore how reasonable this assumption is in depth. We also emphasize that the ruptured aneurysms collected for this work
had a contained rupture, which makes this assumption less critical.

\section{Methods}

Most traditional methods used in machine learning \citep{Hast:Tibs:Frie:2001} for classification assume that:

\begin{itemize}
\item The cost (loss) of classifying one patient as emergent when he is from the elective group
and the cost of classifying one patient as elective when he is from the emergent group are both
the same.
\item The observations in a sample are independent and identically distributed (iid).
\end{itemize}

In the present problem, none of these assumptions are reasonable.
The loss incurred when an emergent patient is misclassified is greater than
the one when this patient is from the elective group.

Furthermore, this study was performed in a case-control scheme (retrospective study): the number of patients in each
group (44 and 100) was fixed prior to data collection.  
Using data from past years from the Western Pennsylvania Allegheny Health System (Table \ref{proportions}),
one can see that these proportions are quite different from what is expected in the target population. In fact, one can
see that the proportions of emergent patients is about 11.4\% per year for all the years presented in the table, rather than the expected percentage of 30.6\% in the sample.
Hence, the iid assumption is also not reasonable.

\begin{table}[H]
\label{proportions}
\begin{center}
\caption{Number of patients with abdominal aortic aneurysm in the Western Pennsylvania Allegheny Health System}
\begin{tabular}{|l||c|c|c|c|c||c|} \hline
&2005&2006&2007&2008&2009&Total\\ \hline
Elective&131&138&141&114&125&649\\ \hline
Emergent&20&15&16&18&15&84\\\hline
\end{tabular}
\end{center}
\end{table}

The two questions that arise when dealing with this nonstandard scenario are:

\begin{itemize}
\item How can we evaluate if a given (fit of a) model is good, i.e., how do we define and estimate an appropriate
risk function?
\item	How can we produce models that give good results? 
\end{itemize}

As we will see, the answer to the first question is to use a modified version of the traditional cross-validation score.
The second question can be resolved by adapting a traditional method in machine learning to this scenario, namely logistic
regression. The use of logistic regressions was also motivated because their results are easy to interpret.
In this work, the need of interpretability of the results is undeniable. 

In the discussion that follows, we will make use of the following notations:

\begin{itemize}
\item $p_0$: probability in the target population of a patient being from the Elective group
\item $p_1$: probability in the target population of a patient being from the Emergent group
\item $l_0$: cost (loss) of classifying a patient from Elective group as being from Emergent group
\item $l_1$: cost (loss) of classifying a patient from Emergent group as being from Elective group
\item $Y$: group of the patient (0 for Elective and 1 for Emergent)
\item $X$: explanatory variables (features)
\end{itemize}

Given a classifier $h:X \rightarrow Y,$ the expected loss function (risk) for a new observation $X$ is
$$ R(h) =  l_1 P(h(X)=0,Y=1) + l_0 P(h(X)=1,Y=0) = l_1 p_1 P(h(X)=0|Y=1)  + l_0 p_0P(h(X)=1|Y=0).$$
In order to estimate $R$, we randomly split the data set in 12 folds with 12 observations each.
For each individual $i$, let 
$$\hat{Y}_i = \hat{h}_{-i}(X_i),$$
where $\hat{h}_{-i}$ is the fitted classifier using all observations but the ones in the fold for which $i$ belongs to.
In this work, we refer to $\hat{Y}_i$ as the cross-validated label for observation $i$.

The estimate of the risk of a fit of classifier $\hat{h}$ we use here is given by
$$ \hat{R}(\hat{h}) = l_1 \hat{p}_1 \hat{P}(\hat{h}(X)=0|Y=1)  + l_0 \hat{p}_0\hat{P}(\hat{h}(X)=1|Y=0),$$
where 
$$\hat{P}(\hat{h}(X)=0|Y=1) = \frac{\sum_{i: Y_i=1} I(\hat{Y}_i=0)}{\sum_{i: Y_i=1} 1}\mbox{, }
\hat{P}(\hat{h}(X)=1|Y=0) = \frac{\sum_{i: Y_i=0} I(\hat{Y}_i=1)}{\sum_{i: Y_i=0} 1},$$

and $\hat{p}_0$ and $\hat{p}_1$ a estimates of $p_0$ and $p_1,$ respectively.

Let $P$ denote the real probability measure, and $P_N$ denote the probability measure
under the "case-control scheme". We assume that $P(x|y) = P_N(x|y),$ that is, the conditional distribution
of the covariates, conditional on the group of the patients, are the same in both sample schemes. This condition is
known as prior probability shift \citep{Quionero-Candela:2009:DSM:1462129}. 

The Bayes classifier classifies a patient as 
Emergent if, and only if,

\begin{align*}
&P(Y=1|x)l_1 > P(Y=0|x)l_0 \iff \frac{P(Y=1|x)}{P(Y=0|x)} > \frac{l_0}{l_1} \iff \frac{P(x|Y=1)P(Y=1)}{P(x|Y=0)P(Y=0)} > \frac{l_0}{l_1} \iff \\[1em]
&\frac{P(x|Y=1)}{P(x|Y=0)} > \frac{l_0 P(Y=0)}{l_1 P(Y=1)} \iff \frac{P(x|Y=1)P_N(Y=1)}{P(x|Y=0)P_N(Y=0)} > \frac{l_0 P(Y=0) P_N(Y=1)}{l_1 P(Y=1) P_N(Y=0)} \iff \\[1em]
& \frac{P_N(Y=1|x)}{P_N(Y=0|x)} > \frac{l_0 P(Y=0) P_N(Y=1)}{l_1 P(Y=1) P_N(Y=0)} \iff P_N(Y=1|x) > \frac{\frac{l_0}{l_1 a}}{1+\frac{l_0}{l_1 a}},
\end{align*}

where $a=\frac{P_N(Y=0)P(Y=1)}{P_N(Y=1)P(Y=0)}.$ The term $P_N(Y=1|x)$ can be estimated using a procedure such as a 
logistic regression, and the right side can be estimated using the data from the Hospital. In this way, we have a plug-in classifier for this
classification problem. This derivation can also be found in \citep{Lin00supportvector}.

Proceeding in an analogous way, we can estimate $P(Y=1|x):$

$$\frac{P(Y=1|x)}{P(Y=0|x)} =  \frac{P_N(Y=1|x)}{P_N(Y=0|x)} \frac{P_N(Y=0)P(Y=1)}{P_N(Y=1)P(Y=0)}.$$

Hence we have

$$P(Y=1|x) = \frac{a\frac{P_N(Y=1|x)}{P_N(Y=0|x)}}{1+a\frac{P_N(Y=1|x)}{P_N(Y=0|x)}},$$

in which the right term can be estimated by plugin of the results of a logistic regression. This derivation is also helpful in 
the case of other plugin classifiers, not only logistic regressions models.

Whenever necessary here, we used the value $p_1=84/733.$ This value was chosen based on the stability
shown in Table \ref{proportions} between the years. We also decided to use the values
$l_0=1$ and $l_1=7.72$ for the loss functions. This was motivated because it is worse to misclassify an emergent
patient than an elective one, as discussed previously. Moreover, for this choice, the risk is proportional to $P(h(X)=0|Y=1)  + P(h(X)=1|Y=0),$ 
that is, it is just the sum of the conditional probabilities of mistake.
The optimal cutoff for the Bayes classifier ($\frac{\frac{l_0}{l_1 a}}{1+\frac{l_0}{l_1 a}}$) in this case is 0.306.

Most of the models we fitted were based on logistic regressions. Rather than using the cutoff derived, 
we decided to use the cutoff as a tuning parameter: for each model, we chose the threshold that
minimizes the empirical risk. It is worth noting that, in fact, this value was in general close to the one
previously derived, as it is going to be shown in the results. Some non-logistic methods were used in order to check the robustness of the models.

\section{Results}

\begin{figure}[H]
\centering
\begin{tabular}{cc}
\includegraphics[scale=0.45]{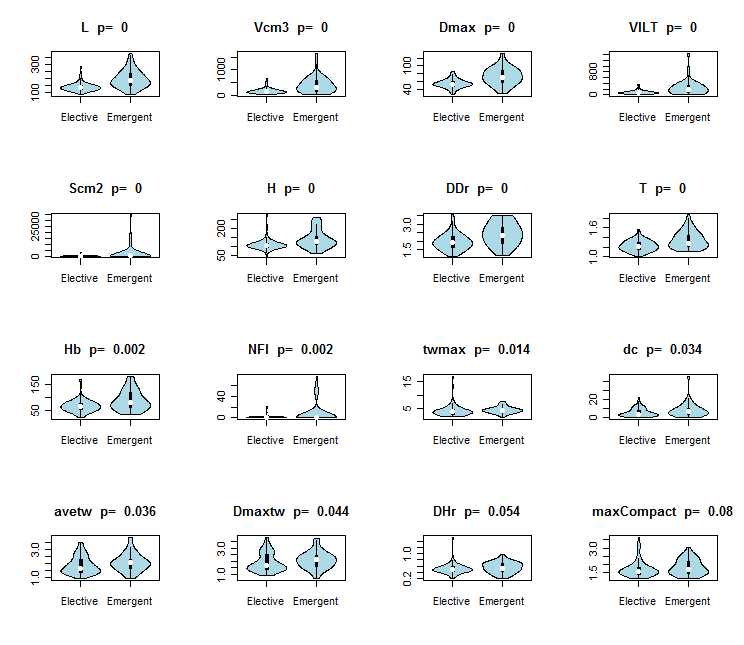}  
\end{tabular}
\caption{Violin Plots for the variables derived from the computed tomography}
\label{violin0}
\end{figure}

\begin{figure}[H]
\centering
\begin{tabular}{cc}
\includegraphics[scale=0.45]{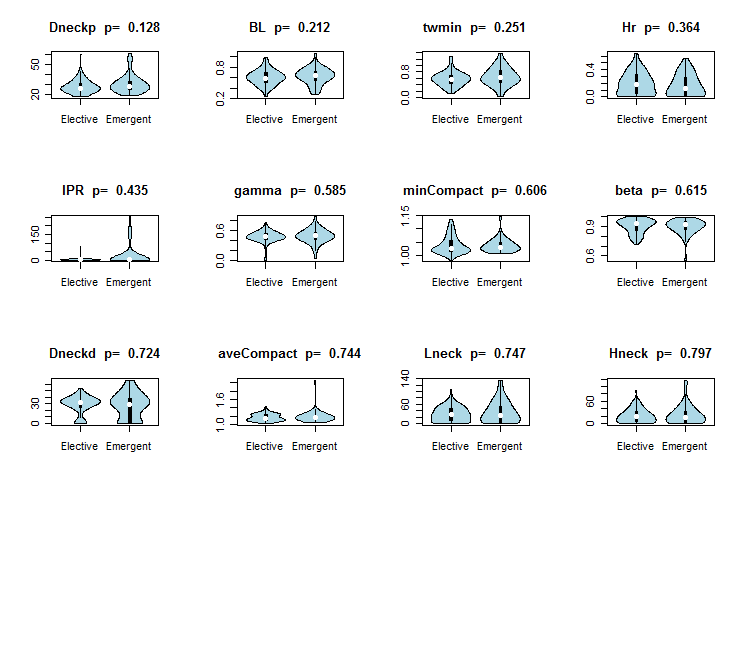}  \\
\end{tabular}
\caption{Violin Plots for the variables derived from the computed tomography}
\label{violin}
\end{figure}

In order to describe the statistical variations of the covariates described in the Appendix, we plotted the violinplots \citep{violin} for each one of them.
Violinplot are boxplots that also contain an estimate of the density of the covariate in each of the groups.
Figures \ref{violin0} and \ref{violin} present the violinplots for the covariates. They are sorted by the p values
obtained in a Mann-Whitney test, which are also presented.

\begin{figure}[h!]
\centering
\begin{tabular}{c}
\includegraphics[scale=0.62]{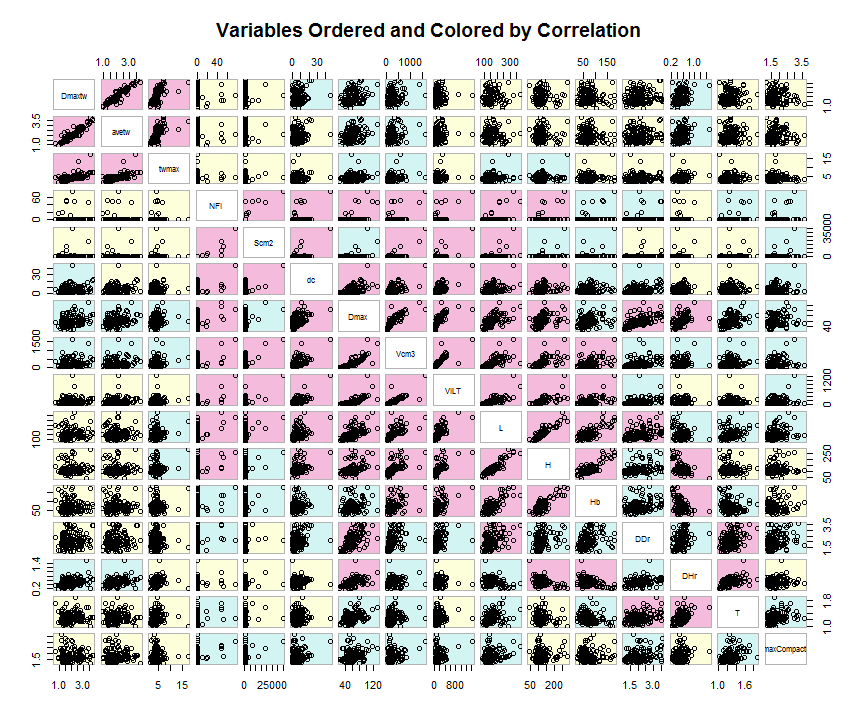}  
\end{tabular}
\caption{Scatter plots of the $16^{th}$ univariately most significant variables}
\label{cor}
\end{figure}

In Figures \ref{violin0} and \ref{violin}, we see that many variables are important predictors of high risk of rupture when
considered separately, especially measures related to the aneurysm's size (such as length, volume and maximum diameter). In order to understand the relationship between them, Figure \ref{cor} shows the scatter plots of the 16 most significant variables on the Mann-Whitney test that was presented.
 
We see that many of the covariates are very correlated to each other.
In what follows we try to consider them jointly by using different models for classification. The following models were implemented (for a reference about these models, see
\citep{Hast:Tibs:Frie:2001}):

\begin{itemize}
\item (Sparse) $L_1$ Regularized Logistic Regression (that is, choosing the coefficients that maximize $\sum y_i(\beta_0+\beta^t x_i)-\log{(1+\exp{(\beta_0+\beta^t x_i}))}-\lambda\sum |\beta_i|$).
\item (SparseL) Maximum Likelihood Logistic Regression with the variables with nonzero coefficients
in the Sparse model
\item (PC) Maximum Likelihood Logistic Regression with the first 5 Principal Components 
of the variables
\item (PCGAMs) Logistic Additive Model with the first 5 Principal Components 
of the variables, using smoothing splines (that is, the mean is modeled as $\log{\frac{\mu(X)}{1-\mu(X)}}=\alpha+s(P_1)+\ldots+s(P_5)$)
\item (PCGAMl) Logistic Additive Model with the first 5 Principal Components 
of the variables, using local polynomials ($1^{st}$ order) (that is, the mean is modeled as $\log{\frac{\mu(X)}{1-\mu(X)}}=\alpha+l(P_1)+\ldots+l(P_5)$)
\item (AIC) Maximum Likelihood Logistic Regression with AIC criterion and stepwise selection

(AIC is $2\frac{\sum y_i(\beta_0+\beta^t x_i)-\log{(1+\exp{(\beta_0+\beta^t x_i}))}}{N}-2\frac{d}{N}$, where $p$ is the number of covariates of the model and $N$ is the sample size)
\item (BIC) Maximum Likelihood Logistic Regression with BIC criterion and stepwise selection (BIC is $\sum y_i(\beta_0+\beta^t x_i)-\log{(1+\exp{(\beta_0+\beta^t x_i}))}-(\log{N})d$)
\item (AICPC) Maximum Likelihood Logistic Regression on the Principal Components with AIC criterion and stepwise selection
\item (RF) Random Forest (which is an average of de-correlated trees)
\end{itemize}

\begin{figure}[H]
\centering
\begin{tabular}{ccc}
\includegraphics[scale=0.27]{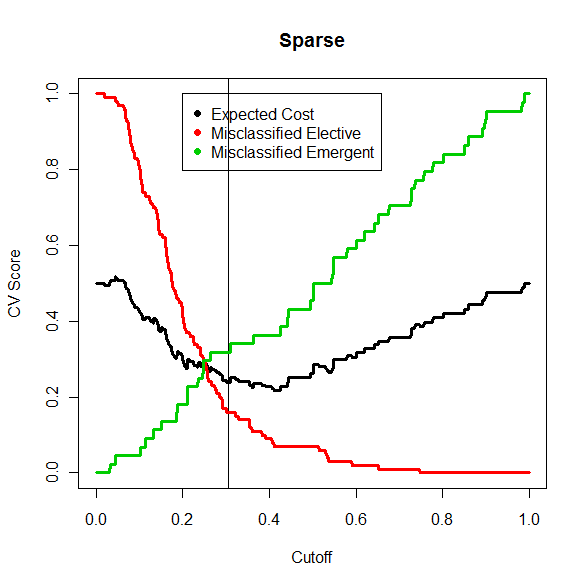}  &
\includegraphics[scale=0.27]{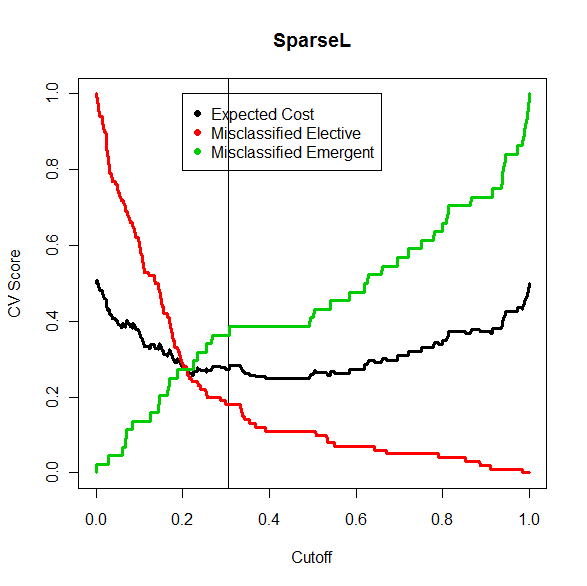}  &
\includegraphics[scale=0.27]{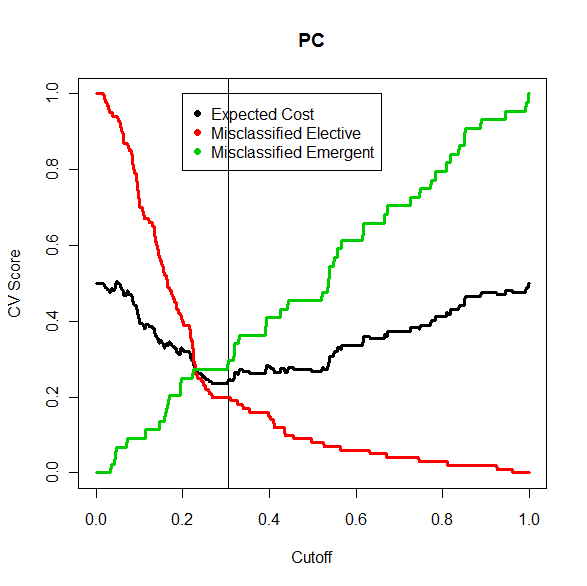}  \\
\includegraphics[scale=0.27]{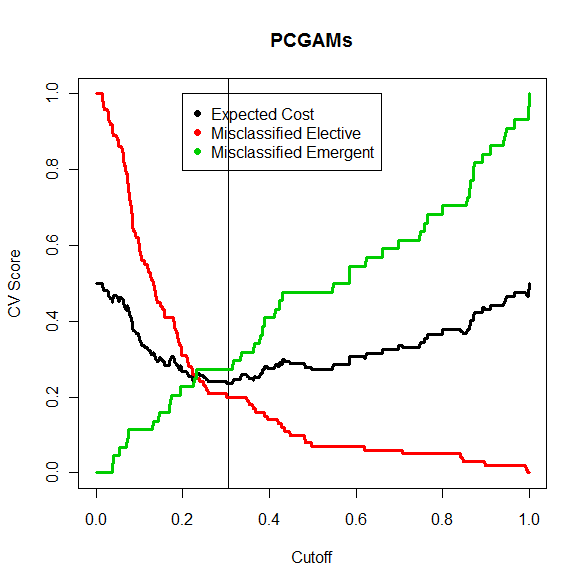} &
\includegraphics[scale=0.27]{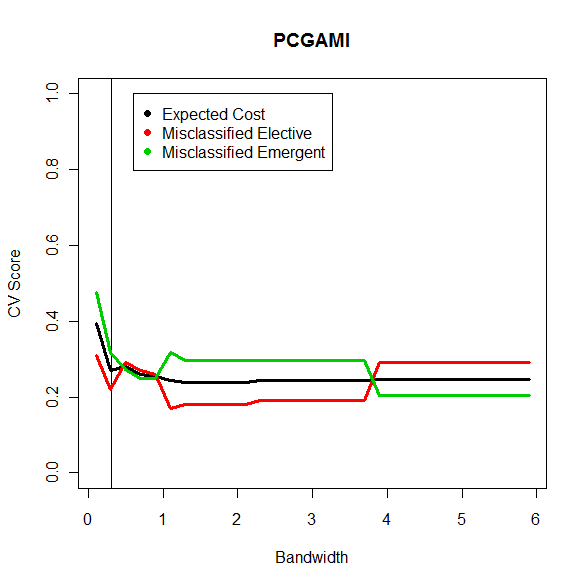}  &
\includegraphics[scale=0.27]{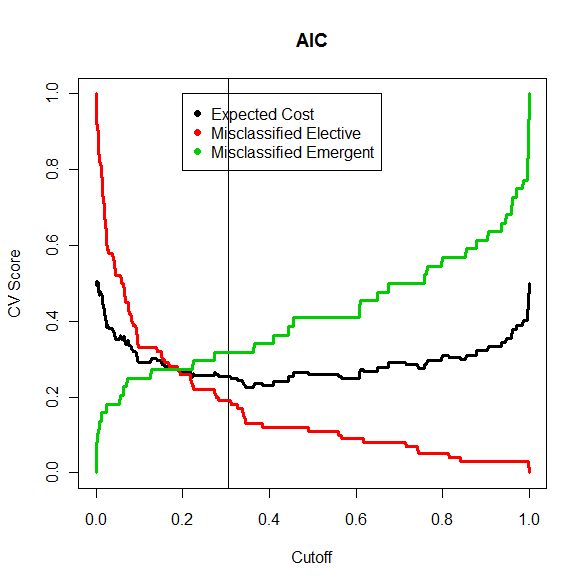}  \\
\includegraphics[scale=0.27]{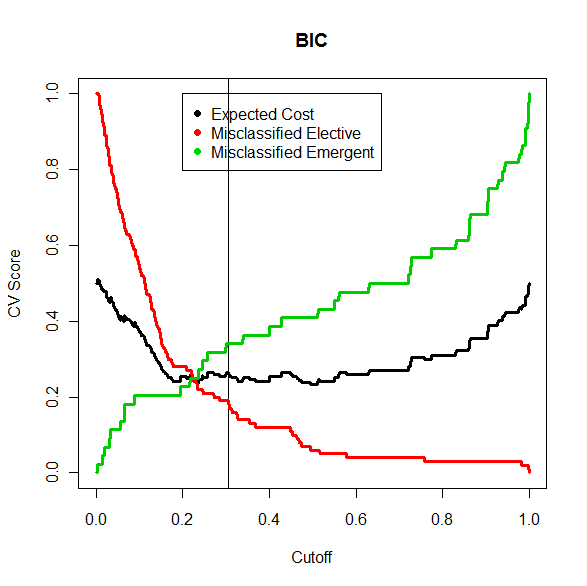}  &
\includegraphics[scale=0.27]{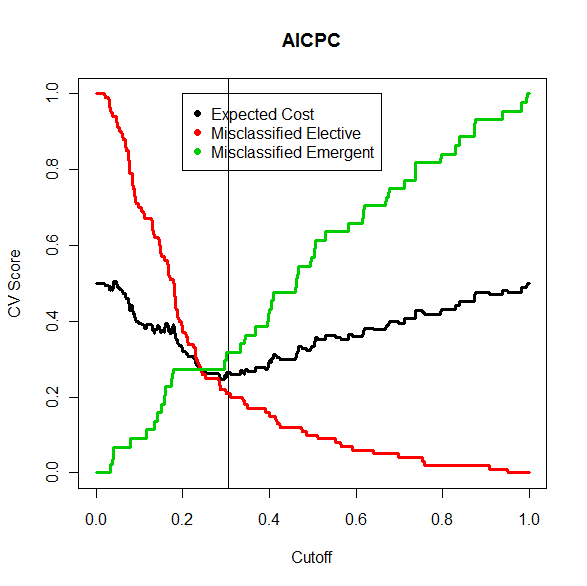} &
\includegraphics[scale=0.27]{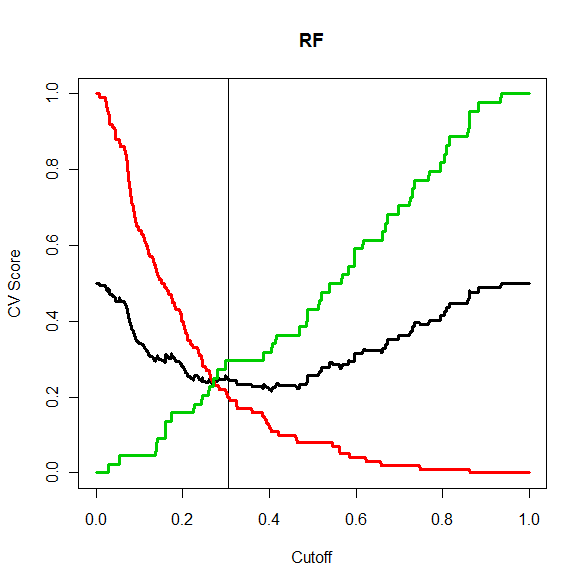}
\end{tabular}
\caption{Results for the estimates of the risk/conditional probabilities for different models and different values of the cutoff on the logistic
regression (and bandwidth for PCGAMl, see text for details). The vertical line represents the optimal cutoff according
to Bayes' rule.}
\label{ROCs}
\end{figure}

Figure \ref{ROCs} shows the (cross-validated) estimates of the conditional misclassification probabilities as
well as the estimate of the risk (expected cost) for the different models and different
values of the cutoffs, as discussed in the previous section. Note that for the PCGAMl model, we have the bandwidth shown on the
x axis rather than the cutoff for the logistic regression. For each value of bandwidth, we chose the optimal cutoff.
Also, for the RF model, note that the same derivation from the previous section applies, once this is also a plugin estimator.
For the sake of comparison, a vertical line is always displayed at the cutoff 0.306, the optimal cutoff derived
on the previous section.
The cross-validated labels were obtained by using a
random 12-fold cross-validation \citep{Hast:Tibs:Frie:2001}. 

\begin{figure}[H]
\centering
\begin{tabular}{cc}
\includegraphics[scale=0.39]{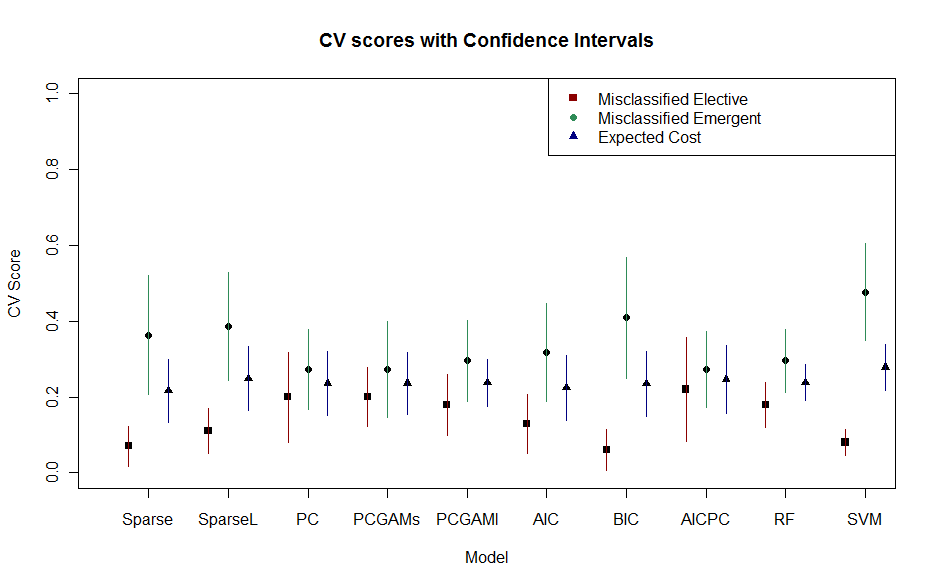}  
\end{tabular}
\caption{90\% bootstrap confidence intervals for the risk and conditional probabilities of mistakes in each model}
\label{Boot}
\end{figure}

In order to evaluate the accuracy of the cross-validation results presented in Figure \ref{ROCs}, 
we plotted the bootstrap 90\% confidence intervals for each of the scores in each of the models for the optimal
cutoffs.
These results are presented in Figure \ref{Boot}. This Figure also contains another model, Support Vector Machine (SVM),
which was used with the purpose of checking the robustness of our conclusions.

We see that the performances of the models are very similar, except SVM. This is expected because no special adaptations
were created to deal with different loss functions and different probabilities. We also notice that the confidence intervals are,
in general, very large, indicating a certain instability of the fitted models. Table \ref{MissResultsT} shows the scores for the models.

\begin{table}[H]
\label{MissResultsT}
\begin{center}
\caption{Summaries (conditional probabilities and risk) of the fitted models.}
\begin{tabular}{|l||c|c|c|c|} \hline
\multicolumn{1}{|l||}{Model}&\multicolumn{1}{c|}{Missclassified Elective}&\multicolumn{1}{c|}{Missclassified Emergent}&\multicolumn{1}{c|}{Expected Cost}\\ \hline
Sparse~&0.070~~~&0.364~~&0.217~~\\ 
SparseL&0.110~~~&0.386~~&0.248~~\\ 
PC~~~~~&0.200~~~~&0.273~~&0.236~~\\ 
PCGAMs~&0.200~~~~&0.273~~&0.236~~\\ 
PCGAMl~&0.180~~~&0.295~~&0.238~~\\ 
AIC~~~~&0.130~~~&0.318~~&0.224~~\\ 
BIC~~~~&0.060~~~&0.409~~&0.235~~\\ 
AICPC~~&0.220~~~&0.273~~&0.246~~\\ 
RF~~~~~&0.180~~~&0.295~~&0.238~~\\ 
SVM~~~~&0.090~~~&0.410~~&0.250~~\\ 
\hline
\end{tabular}
\end{center}
\end{table}

\begin{figure}[H]
\centering
\begin{tabular}{cc}
\includegraphics[scale=0.45]{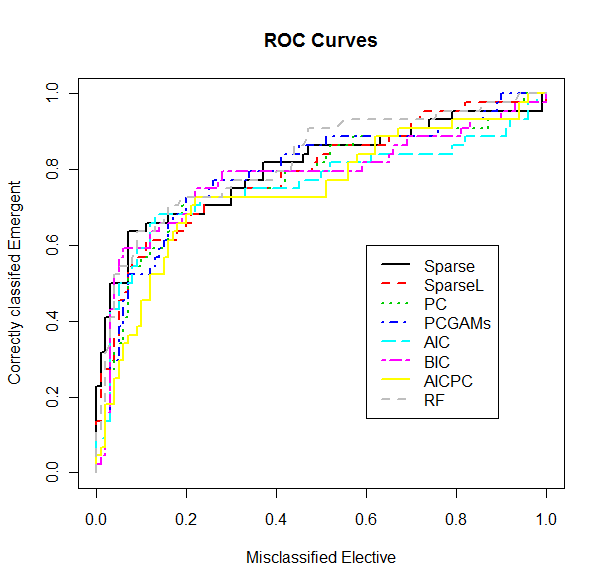}  
\end{tabular}
\caption{ROC curves for the models with continuous output (with the cross-validated labels).}
\label{ROC}
\end{figure}

Figure \ref{ROC} presents the ROC curves (for different cutoff values) of the fitted models which have a continuous output. Again, one can
see that all of the models are very similar. In this figure, we also used cross-validated labels.

\begin{figure}[H]
\centering
\begin{tabular}{cc}
\includegraphics[scale=0.38]{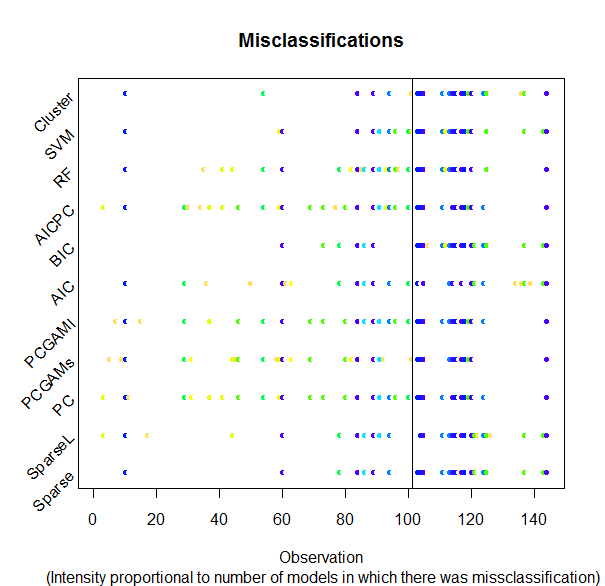}  
\end{tabular}
\caption{Cross-validated misclassifications in each models. Color intensity is proportional to the number of  misclassifications  
for the patient. First 100 patients are elective, last 44 are emergent.}
\label{miss}
\end{figure}

Although the estimated risks of the models are very similar, we can ask ourselves if, in each of the models, we are
misclassifying the same observations or not. Figure \ref{miss} shows, for each observation and each model,
a dot in case this observation was misclassified on that model. The intensity of the color is proportional
to the number of models in which the observation was misclassified. We also added a 2-means cluster analysis, using
all covariates.
Note that the observations to the right of the vertical line are patients from the emergent group, while
the ones to the left are the ones from the elective group. The labels used were again the cross-validated ones.

We can see that most misclassified patients have dots with dark colors. This indicates that most misclassified patients are being misclassified
in all fitted models (even on the cluster analysis, which doesn't use the labels to train the classifier). In future work, we intend to investigate which characteristics are shared by these patients.

\begin{figure}[H]
\centering
\begin{tabular}{cc}
\includegraphics[scale=0.35]{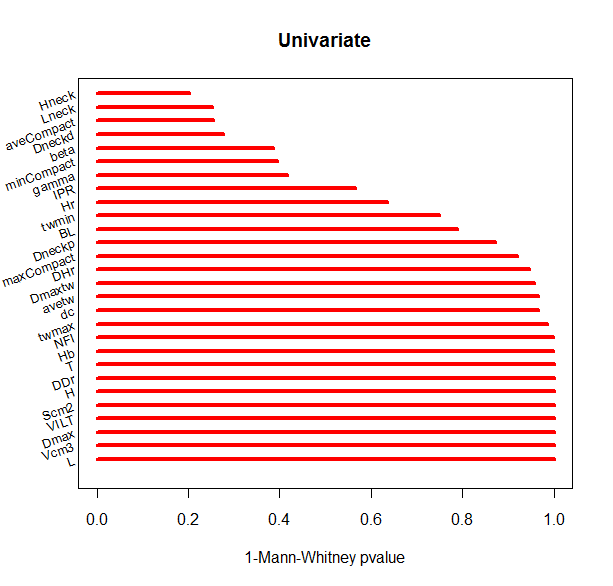}  &
\includegraphics[scale=0.35]{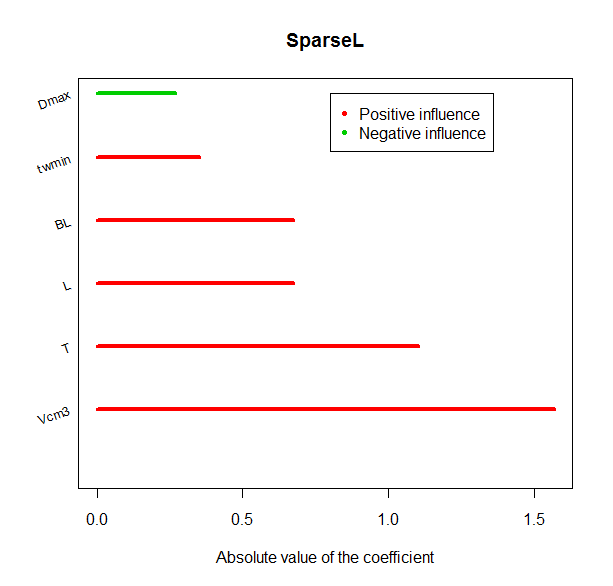} \\ 
\includegraphics[scale=0.35]{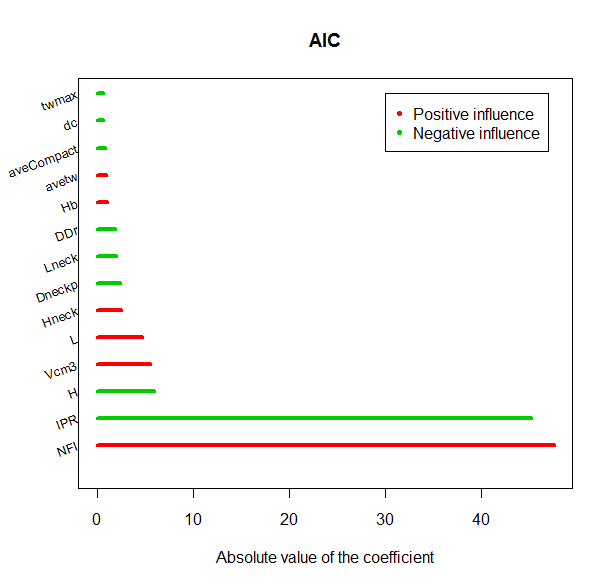}  &
\includegraphics[scale=0.35]{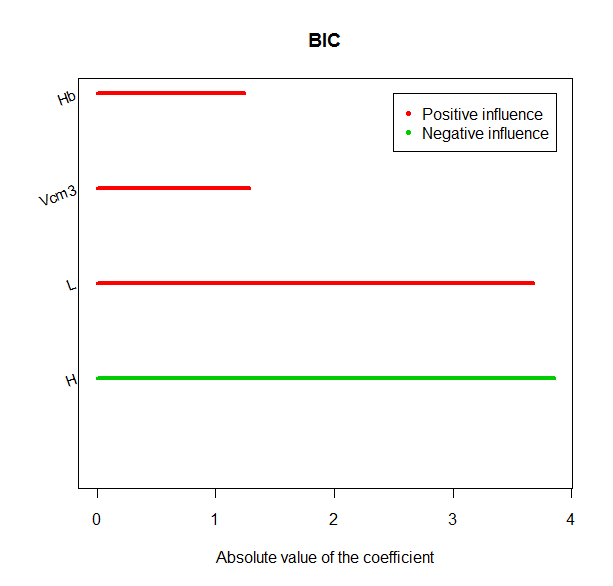}  \\
\end{tabular}
\caption{Variable importance for univariate Mann-Whitney tests, SparseL, AIC and BIC, respectively.}
\label{Importance0}
\end{figure}

\begin{figure}[H]
\centering
\begin{tabular}{cc}
\includegraphics[scale=0.35]{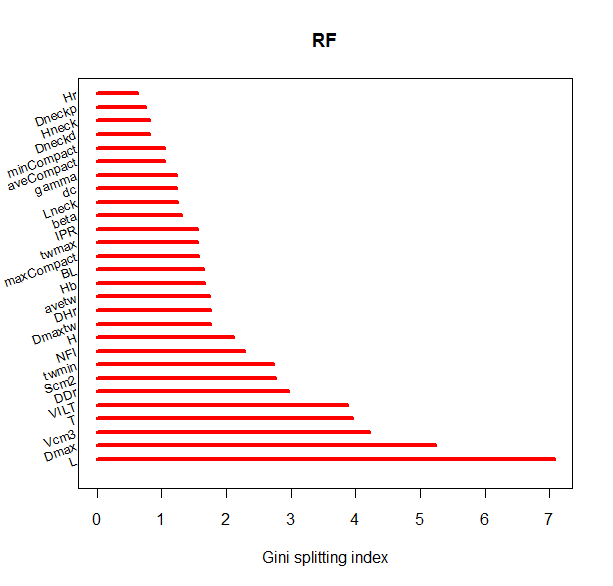}  
\end{tabular}
\caption{Variable importance for RF model.}
\label{Importance}
\end{figure}

Figures \ref{Importance0} and \ref{Importance} show the importance of each variable according to some of the models presented.
The univariate importance was measured in terms of the Mann-Whitney pvalues, while the importance of the logistic models
was measured in terms of the coefficients of the covariates after they were standardized.
Also, for the Random Forest model, we calculated the importance of each variable based on the Gini splitting index \citep{Hast:Tibs:Frie:2001}. The greater the Gini index, the greater the
variable importance is. In the graphics associated to maximum likelihood logistic regression, a red line represents
that the chance that the patient is from the emergent group increases as the covariate increases (conditional
on the others). On the other hand,
a green line indicates
that the chance that the patient is from the emergent group decreases as the covariate increases.

%
%

It is interesting to note that some of these variables, like twmin, are not important univariately.
Also, many of the important variables are related to the size of the aneurysm (as for instance L, H and Vcm3). The summaries of
Wall Thickness that were considered to be important by this models were the minimum and the maximum. Tortuosity is one
variable that is not directly related to the size of the aneurysm but seems to be very important.

\section{Conclusions and future work}

In this work, we see that the introduction of a different loss functions (that reflect that it is
worse to misclassify patients from the Emergent group) and information regarding prior
probabilities of each group on the target population allow us to adapt the traditional logistic 
regression to a new context. Moreover, it leads us to the idea of using the cutoff of the logistic regression
as a tuning parameter. The same derivations work for other plugin methods.

We saw that the confidence intervals for the risks were large in the fitted models. This leads us to believe that the use of
even more complex models would not be helpful for current sample sizes. 

The fitted models suggest that with the covariates used we are able to do a fair work in classifying patients from
the elective group, however more informative features would be desirable to classify correctly patients from
emergent group. It is worth noting that the inclusion of interactions in the models did not improve the estimated
risks (though these results were omitted in this paper). All fitted models had a similar performance both in terms
of the final estimate of their risk and also in terms of which patients were being misclassified.

Possible further investigations for this work include:

\begin{itemize}
\item Among the patients from Emergent group, compare the ones that were in pain when the computed tomography was
taken and those who already had the aneurysm ruptured at that time. These comparisons would be useful to try to validate the assumption that
the rupture of the aneurysm does not change the covariates used in this work.
\item Try to understand what common characteristics patients that were misclassified share. Sociodemographic variables may help.
\item Check whether the probabilities on the logistic regressions from misclassified individuals were close to the cutoff.
\item Introduce a measure of confidence for the classifications, e.g. the brier score $\sum_i (y_i - \hat{y}_i)^2,$
where $y_i$ denotes the category of observation $i$ and $\hat{y}_i$ denotes the fitted value (score) for observation $i$.
\item Focus on trying to find new good predictors rather than trying to fit complex models. Combining information on wall thickness with information on wall stress at the same location may yield good results.
\item Increase the number of samples, especially from the Emergent group.
\item Investigate the relationship between wall stress (still to be measured) and the geometric features used here.
The hypothesis is that the wall stress is highly associated to the rupture of an aneurysm.
\end{itemize}

\bibliographystyle{plainnat}
\bibliography{paper}

\appendix
\section{Covariates}

1-dimensional indices:
\begin{itemize}
\item Dmax: Maximum diameter
\item Dneckp: Distal neck diameter
\item Dneckd: Proximal neck diameter
\item H: Height of AAA
\item L: Length of AAA centerline
\item Hneck: Height of neck
\item Lneck: Length of neck centerline
\item Hb: Bulge Height
\item dc: Centroid distance of Dmax
\item maxCompact: Maximum Compactness
\item minCompact: Minimum Compactness
\item aveCompact: Average Compactness
\end{itemize}
\vspace{3mm}

2-dimensional indices:
\begin{itemize}
\item DHr: Diameter-Height ratio
\item DDr: Diameter-Diameter ratio
\item Hr: Height ratio
\item BL: Bulge location
\item beta: Asymmetry
\item T: Tortuosity
\end{itemize}
\vspace{3mm}

3-dimensional indices:
\begin{itemize}
\item Vcm3: AAA Volume
\item Scm2: AAA Surface Area
\item VILT: Intraluminal thrombus volume
\item gamma: AAA sac to ILT volume ratio
\item IPR: Isoperimetric Ratio
\item NFI: Non-fusiform Index
\end{itemize}
\vspace{3mm}

Summaries of wall thickness:
\begin{itemize}
\item twmax: Maximum wall thickness
\item twmin: Minimum wall thickness
\item avetw: Avarage wall thickness
\item Dmaxtw: Maximum wall thickness at Dmax
\end{itemize}

\begin{figure}[H]
\centering
\label{illustration}
\begin{tabular}{cc}
\includegraphics[scale=0.45]{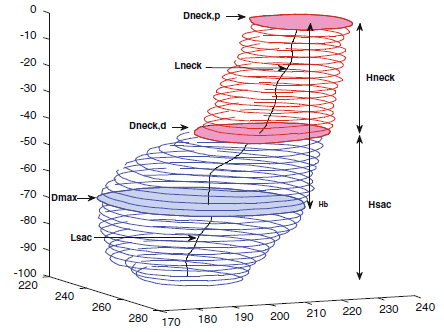}  &
\includegraphics[scale=0.4]{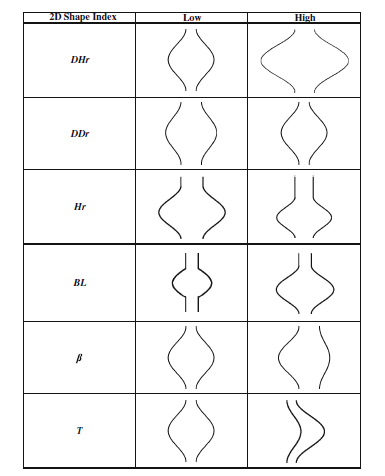} \\ 
\includegraphics[scale=0.4]{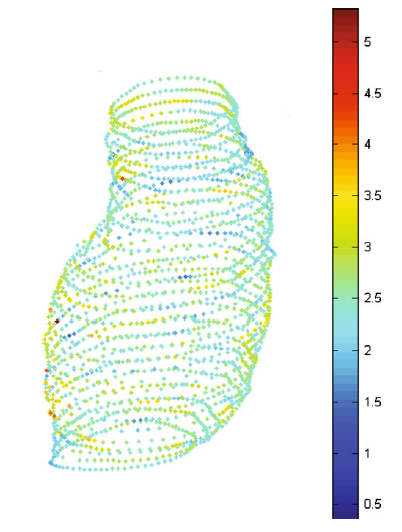}  &
\includegraphics[scale=0.4]{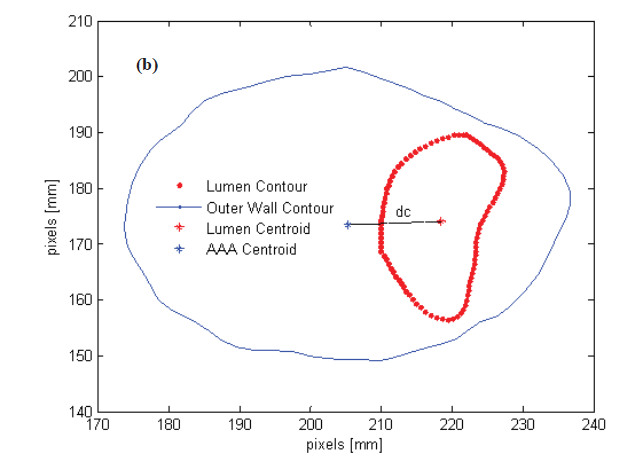}  \\
\end{tabular}
\caption{Visual illustration of some of the measures. Bottomleft represents
the wall stress of one patient at different locations. (Courtesy of Ender A. Finol and Judy Shum)}
\end{figure}

\end{document}